\documentclass[twocolumn,showpacs,preprintnumbers,superscriptaddress,prl,amsmath,amssymb]{revtex4}

\usepackage{graphicx}
\usepackage{dcolumn}
\usepackage{bm}

\begin{document}

\title{Magnetic and structural transitions in La$_{0.4}$Na$_{0.6}$Fe$_2$As$_2$ single crystals}

\author{J.-Q. Yan}
\email{yanj@ornl.gov}
\affiliation{Materials Science and Technology Division, Oak Ridge National Laboratory, Oak Ridge, Tennessee 37831, USA}
\affiliation{Department of Materials Science and Engineering, University of Tennessee, Knoxville, Tennessee 37996, USA}

\author{S. Nandi}
\affiliation{J\"{u}lich Centre for Neutron Science JCNS and Peter Gr\"{u}nberg Institut PGI, JARA-FIT, Forschungszentrum J\"{u}lich GmbH, D-52425 J\"{u}lich, Germany}
\affiliation{J\"{u}lich Centre for Neutron Science JCNS, Forschungszentrum J\"{u}lich GmbH, Outstation at MLZ, Lichtenbergstraße 1, D-85747 Garching, Germany}

\author{B. Saparov}\thanks{Present address: Department of Mechanical Engineering and Materials Science, Duke University, Durham, NC 27708, USA}
\affiliation{Materials Science and Technology Division, Oak Ridge National Laboratory, Oak Ridge, Tennessee 37831, USA}

\author{P. \v{C}erm\'{a}k}
\affiliation{J\"{u}lich Centre for Neutron Science JCNS, Forschungszentrum J\"{u}lich GmbH, Outstation at MLZ, Lichtenbergstraße 1, D-85747 Garching, Germany}

\author{Y. Xiao}
\affiliation{J\"{u}lich Centre for Neutron Science JCNS and Peter Gr\"{u}nberg Institut PGI, JARA-FIT, Forschungszentrum J\"{u}lich GmbH, D-52425 J\"{u}lich, Germany}

\author{Y. Su}
\affiliation{J\"{u}lich Centre for Neutron Science JCNS, Forschungszentrum J\"{u}lich GmbH, Outstation at MLZ, Lichtenbergstraße 1, D-85747 Garching, Germany}

\author{W. T. Jin}
\affiliation{J\"{u}lich Centre for Neutron Science JCNS and Peter Gr\"{u}nberg Institut PGI, JARA-FIT, Forschungszentrum J\"{u}lich GmbH, D-52425 J\"{u}lich, Germany}
\affiliation{J\"{u}lich Centre for Neutron Science JCNS, Forschungszentrum J\"{u}lich GmbH, Outstation at MLZ, Lichtenbergstraße 1, D-85747 Garching, Germany}

\author{A. Schneidewind}
\affiliation{J\"{u}lich Centre for Neutron Science JCNS, Forschungszentrum J\"{u}lich GmbH, Outstation at MLZ, Lichtenbergstraße 1, D-85747 Garching, Germany}

\author{Th. Br\"{u}ckel}
\affiliation{J\"{u}lich Centre for Neutron Science JCNS and Peter Gr\"{u}nberg Institut PGI, JARA-FIT, Forschungszentrum J\"{u}lich GmbH, D-52425 J\"{u}lich, Germany}
\affiliation{J\"{u}lich Centre for Neutron Science JCNS, Forschungszentrum J\"{u}lich GmbH, Outstation at MLZ, Lichtenbergstraße 1, D-85747 Garching, Germany}

\author{R. W. McCallum}
\affiliation{Division of Materials Science and Engineering, Ames Laboratory, Ames, Iowa 50010, USA}
\affiliation{Department of Materials Science and Engineering, Iowa State University, Ames, Iowa 50010, USA}

\author{T. A. Lograsso}
\affiliation{Division of Materials Science and Engineering, Ames Laboratory, Ames, Iowa 50010, USA}
\affiliation{Department of Materials Science and Engineering, Iowa State University, Ames, Iowa 50010, USA}

\author{B. C. Sales}
\affiliation{Materials Science and Technology Division, Oak Ridge National Laboratory, Oak Ridge, Tennessee 37831, USA}

\author{D. G. Mandrus}
\affiliation{Materials Science and Technology Division, Oak Ridge National Laboratory, Oak Ridge, Tennessee 37831, USA}\affiliation{Department of Materials Science and Engineering, University of Tennessee, Knoxville, Tennessee 37996, USA}

\date{\today}

\begin{abstract}
La$_{0.4}$Na$_{0.6}$Fe$_2$As$_2$ single crystals have been grown out of an NaAs flux in an alumina crucible and characterized by measuring magnetic susceptibility, electrical resistivity, specific heat, as well as single crystal x-ray and neutron diffraction. La$_{0.4}$Na$_{0.6}$Fe$_2$As$_2$ single crystals show a structural phase transition from a high temperature tetragonal phase to a low-temperature orthorhombic phase at T$_s$\,=\,125\,K. This structural transition is accompanied by an anomaly in the temperature dependence of electrical resistivity, anisotropic magnetic susceptibility, and specific heat. Concomitant with the structural phase transition, the Fe moments order along the \emph{a} direction with an ordered moment of 0.7(1)\,$\mu_{\textup{B}}$ at \emph{T}\,=\,5 K. The low temperature stripe antiferromagnetic structure is the same as that in other \emph{A}Fe$_{2}$As$_{2}$ (\emph{A}\,=\,Ca, Sr, Ba) compounds. La$_{0.5-x}$Na$_{0.5+x}$Fe$_2$As$_2$ provides a new material platform for the study of iron-based superconductors where the electron-hole asymmetry could be studied by simply varying La/Na ratio.

\end{abstract}


\maketitle

\section{Introduction}

Since the first report of superconductivity with a transition temperature (T$_c$) of 26\,K in LaFeAsO$_{1-x}$F$_x$ in 2008,\cite{Hosono2008} tremendous efforts have been made
to explore new iron-based superconductors. Superconductivity was observed in a wide variety of iron-based materials, such as '122' ThCr$_2$Si$_2$-type \emph{A}Fe$_2$As$_2$ (\emph{A}\,=\,alkaline earth),\cite{Johrendt2008} '111' Cu$_2$Sb-type LiFeAs,\cite{Jin2008} '11' PbO-type FeSe,\cite{Wu2008} '42622' Sr$_2$FeO$_3$CuS-type Sr$_4$Sc$_2$O$_6$Fe$_2$P$_2$,\cite{Ogino2009} and '10-3-8' phase Ca$_{10}$(Pt$_3$As$_8$)(Fe$_2$As$_2$)$_5$.\cite{Ni2011} Among them, the '122' system has attracted the most interest because (1) sizable single crystals could be grown which makes it possible to investigate intrinsic physical properties and the close interplay between structure, magnetism, and superconductivity by various techniques, (2) the materials show rich chemistry and can be both hole- and electron-doped. It thus has been a model platform for unconventional superconductivity.\cite{Mandrus2010}

As shown in the inset of Fig.\,\ref{XRD-1}, ThCr$_2$Si$_2$-type 122 compounds have a layered tetragonal structure at room temperature. The FeAs layers formed by conjugated [FeAs$_4$] tetrahedra are separated by layers of alkaline earth atoms. In the structure, atoms are located at \emph{A} 2\emph{a} (0,\,0,\,0), Fe 4\emph{d} (1/2, 0, 1/4), As 4\emph{e} (0, 0, z). 122 compounds show rich chemistry by allowing substitutions at all three crystallographic sites. For example, alkaline earth elements can be replaced by rare earth ions or alkali ions;\cite{Johrendt2008,XHChen2008, CaPrFe2As2} Fe by various transition metal ions;\cite{PaulReview} and As by P.\cite{GHCao2009} Exotic ground states were obtained with chemical substitution and appropriate substitution at each site can induce superconductivity. As in cuprate superconductors, the electron-hole asymmetry was observed in the phase diagram of iron-based superconductors, which shows the doping dependence of magnetic, structural, and superconducting transitions. The asymmetry is well illustrated in doped BaFe$_2$As$_2$ by comparing hole-doped Ba$_{1-x}$K$_x$Fe$_2$As$_2$ and electron-doped BaFe$_{2-x}$Co$_x$As$_2$, mainly due to the availability of high quality single crystals which enable systematic studies using various probes.\cite{Mandrus2010, Osborn2012} Following the convention in cuprate superconductors, the FeAs layer is sometimes named as conducting layer and the alkaline earth metal networks as spacing layer. In Ba$_{1-x}$K$_x$Fe$_2$As$_2$, K-doping takes place at the spacing layer while FeAs layers remain intact. In contrast, Co substitution in BaFe$_{2-x}$Co$_x$As$_2$ disturbs the contiguity of the [FeAs$_4$] tetrahedra and interferes with superconductivity in the conducting layers. This effect coming from substitution at different crystallographic sites has been suggested to contribute to the electron-hole asymmetry in a phase diagram of iron-based superconductors.\cite{Osborn2012} From materials design point of view, an ideal material for the study of electron-hole asymmetry should meet the following criteria: (1) substitution takes place at only one crystallographic site, (2) the material can be tuned from electron-doped to hole-doped by varying the ratio between the substitutional and substituted atoms, and (3) the substitutional and substituted atoms should have similar size to minimize the steric effect. Unfortunately, none of the presently studied materials meets the above requirements.

In this work, we report the magnetic and structural transitions in La$_{0.4}$Na$_{0.6}$Fe$_2$As$_2$ single crystals. La$_{0.4}$Na$_{0.6}$Fe$_2$As$_2$ shows a structural phase transition from a high temperature tetragonal phase to a low-temperature orthorhombic phase at T$_s$\,=\,125\,K. Concomitant with the structural phase transition, the Fe moments order along the \emph{a} direction with an ordered moment of 0.7(1) \,$\mu_{\textup{B}}$ at \emph{T} = 5 K, with the same stripe antiferromagnetic structure as that in other \emph{A}Fe$_{2}$As$_{2}$ (\emph{A} = Ca, Sr, Ba) compounds. The magnetic and  structural transitions are accompanied by an anomaly in the temperature dependence of electrical resistivity, anisotropic magnetic susceptibility, and specific heat. The results show that La$_{0.5-x}$Na$_{0.5+x}$Fe$_2$As$_2$, with an alkali metal and a rare earth ion at the spacing layer, provides a new material platform for the study of iron-based superconductors. More importantly, from simple electron counting, the material could be tuned from electron-doped (x$<$0) to hole-doped (x$>$0) by varying the ratio between the alkali metal and rare earth ions.

\section{Experimental Details}
La$_{0.4}$Na$_{0.6}$Fe$_2$As$_2$ single crystals were grown by accident when testing suitable crucibles for the growth of LaFeAsO crystals in NaAs flux. NaAs is an appropriate flux for the growth of millimeter-sized \emph{R}FeAsO single crystals. The synthesis of NaAs flux and detailed crystal growth process for \emph{R}FeAsO have been reported previously.\cite{YanAPLGrowth} \cite{YanFluxRequirement} La$_{0.4}$Na$_{0.6}$Fe$_2$As$_2$ single crystals were grown in NaAs flux with a similar procedure. LaAs powder was first prepared by firing La filings and arsenic chunks in a sealed quartz tube at 900$^o$C for 15 hours. LaAs, (1/3Fe\,+\,1/3Fe$_2$O$_3$) or FeO, and NaAs were mixed with the molar ratio of 1:1:15 inside of a dry glove box filled with He. The mixture was then loaded into a 2\,ml alumina crucible covered with an alumina cap. The alumina crucible was further sealed inside of a Ta tube under ~1/3 atmosphere of argon gas. The Ta tube was then sealed in an evacuated quartz tube. The entire assembly was heated inside of a box furnace sitting in a fume hood. The furnace was programmed to heat up to 1150$^o$C over 12 hours, hold at 1150$^o$C for 16 hours, and then cooled to 850$^o$C over 100 hours. Once the furnace reached 850$^o$C, the furnace was turned off. Plate-like single crystals were separated from flux by rinsing them with deionized water in a closed fume hood.

Elemental analysis of the crystals was performed using both wavelength dispersive x-ray spectroscopy (WDS) in the electron probe microanalyzer of a JEOL JXA-8200 electron microprobe and energy-dispersive x-ray spectroscopy using a Hitachi-TM3000 microscope equipped with a Bruker Quantax 70 EDS system. Both techniques confirmed the atomic ratio La:Fe:As=0.4:2:2, but the determined Na content varies. The variation of Na content might result from the followings: (1) contamination from the flux. When washing crystals out of flux in water, NaAs reacts with water forming sodium arsenate hydrate staying on the surface. This might contaminate some measurements; (2) reaction with water. Since the sample is moisture sensitive, some Na in the surface layers might be removed by reacting with water when isolating the crystals from flux, or moisture when exposing in air; and (3) the alkali ions tend to be nonuniform in crystals. To better determine the composition, we utilized single crystal x-ray diffraction. Instead of washing the crystals out of NaAs flux, one small piece of crystal was manually picked up from the flux inside of a dry glove box. The crystal was then cut to a suitable size (0.1\,mm on all sides) inside Paratone N oil under an optical microscope. Because the crystals are very soft and malleable, extreme care was taken not to deform them. Single crystal x-ray diffraction measurements were performed on a Bruker SMART APEX CCD-based single crystal X-ray diffractometer with Mo K$_\alpha$ ($\lambda$ = 0.71073\,\emph{${\AA}$}) radiation. Data collection was performed at  100(2)\,K and 173(2)\,K, respectively. The sample was cooled using a cold nitrogen stream. The structure solution by direct methods and refinement by full matrix least-squares methods on \emph{F}$^2$ were carried out using the SHELXTL software package. SADABS was used to apply absorption correction.

Room temperature x-ray diffraction patterns were collected on a X'Pert PRO MPD X-ray Powder Diffractometer using the Ni-filtered Cu-K$_\alpha$ radiation. Magnetic properties were measured with a Quantum Design (QD) Magnetic Property Measurement System in the temperature interval 1.8\,K\,$\leq$\,T\,$\leq$\,350\,K. The temperature dependent specific heat and electrical transport data were collected using a 14 Tesla QD Physical Property Measurement System in the temperature range of 1.9\,K\,$\leq$\,T\,$\leq$\,300\,K. The electrical resistivity measurements were performed with a four-probe technique on specimens having typical size of 0.5$\times$0.01$\times$2 mm$^3$. Four pieces of platinum wire are attached to the surface of the specimen via silver paste.

For the neutron scattering measurements, a
3\,mg as-grown square shaped single crystal of approximate dimensions
2.5\,mm$\times$2.5\,mm$\times0.1$\,mm was selected. High-resolution
elastic neutron scattering experiments were carried out on the cold-neutron
triple-axis spectrometer PANDA at the MLZ in Garching (Germany). Vertically
focused Pyrolytic Graphite (PG) (0 0 2) monochromator and analyzer
were used. The measurements were carried out with fixed incoming and
final wave vectors of k$_{i}$\,=\,k$_{f}$\,=\,1.57\,${\AA}$$^{-1}$ and 2.57\,${\AA}$$^{-1}$,
which correspond to neutron wavelengths of 4.002\,${\AA}$ and 2.445\,${\AA}$, respectively.
A liquid nitrogen cooled Be filter after the sample was used to reduce the second order contamination for the k$_{f}$\,=\,1.57\,${\AA}$$^{-1}$. For the shorter wavelengths (k$_{f}$\,=\,2.57\,${\AA}$$^{-1}$), PG filter was used on the same place. Due to the geometrical limitation, the shorter wavelength
was employed for the measurements of the integrated intensities of
the structural and magnetic peaks. The larger wavelength was used
for the rest of the measurements due to the higher flux. The single crystal
was mounted on an Aluminum pin using very small amount of GE-Varnish
and mounted inside a bottom-loading closed cycle refrigerator. Slits
before and after the sample were used to reduce the background. The
measurements were performed in the (1 1 0)$_{T}$-(0 0 1)$_{T}$ scattering
plane. Measurements at PANDA were performed at temperatures between
5 and 150\,K. We will use tetragonal (\emph{T}) and orthorhombic
(\emph{O}) subscripts for the reflections whenever necessary.

\section{Results and Discussion}
\begin{figure}
\centering \includegraphics[width = 0.48\textwidth] {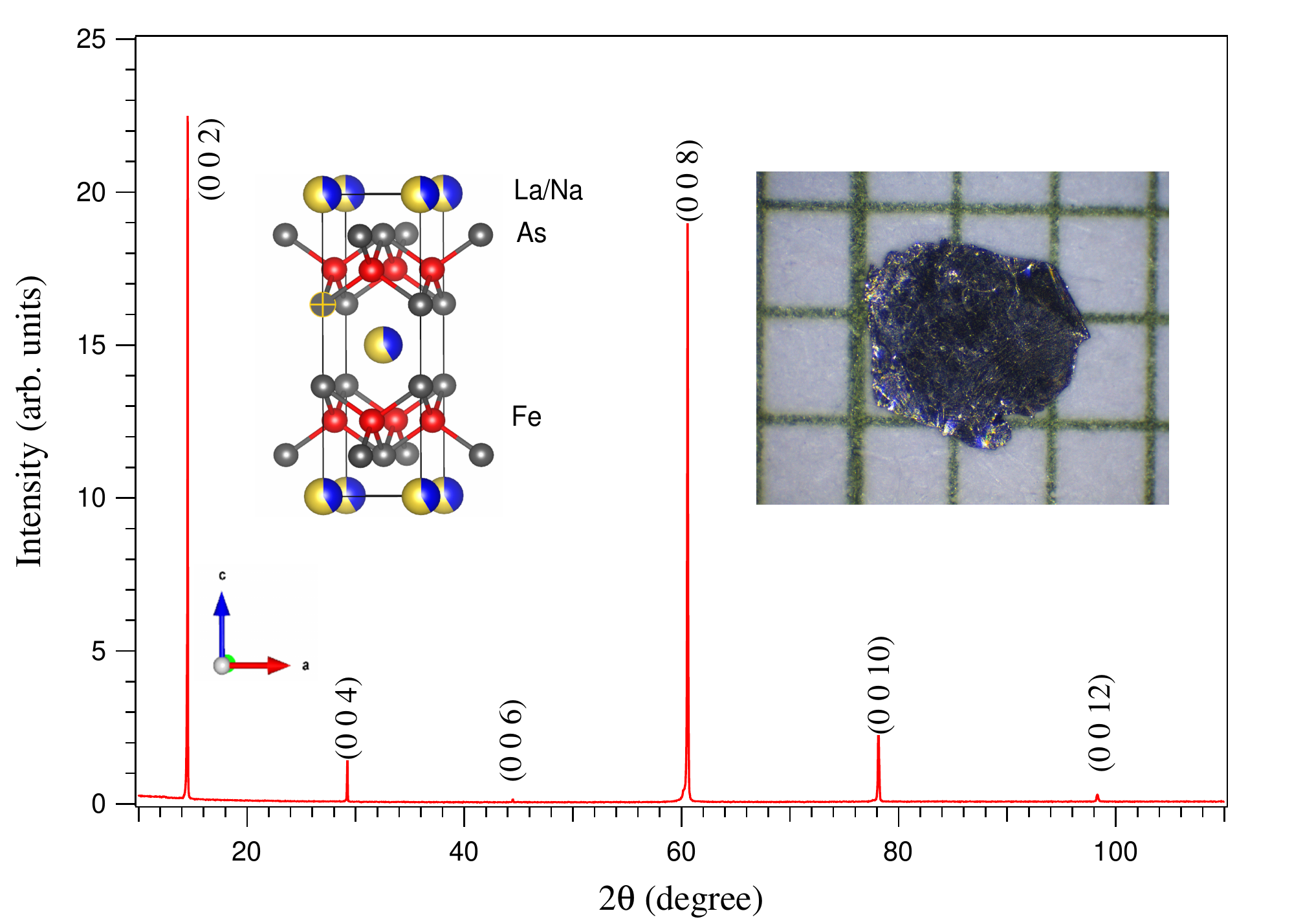}
\caption{(color online) (0 0 \emph{l}) reflections from one piece of plate-like crystal. The left inset shows the structure of the high temperature tetragonal phase. The right inset shows the photograph of a single crystal on a mm grid. The crystallographic \emph{c} axis is perpendicular to the plane of the plate.} \label{XRD-1}
\end{figure}

The inset of Fig.\,\ref{XRD-1} shows a picture of one piece of single crystal against a millimeter scale. The as-grown crystals are platelike with typical dimensions of 2$\times$2$\times$0.1\,mm$^3$. We tried x-ray powder diffraction measurement of pulverized single crystals, but the powder has strong preferred orientation and impurities are always observed because the ground powder is moisture sensitive. We thus examined the \emph{ab}-plane of the crystals by mounting the platelike crystals on a powder diffractometor. As expected, only (0\,0\,\emph{l}) (\emph{l}\,=\,2n) reflections were observed (see Fig.\,\ref{XRD-1}), indicating the crystallographic \emph{c}-axis is perpendicular to the plane.

The \emph{c}-lattice parameter of $\sim$12.20${\AA}$ obtained from 2$\theta$ values of (0\,0\,\emph{l}) reflections is much larger than 8.746${\AA}$ expected for LaFeAsO.\cite{YanAPLGrowth} Elemental analysis confirmed Fe/As$\approx$1 in the as-grown crystals, but the results show (1) Fe/La ratio is $\approx$5 which is significantly different from 1 expected for LaFeAsO, and (2) a significant amount of Na was observed while previous studies observed no Na in LaFeAsO crystals grown in a similar procedure. The above suggests that the obtained crystals are not LaFeAsO. Considering the \emph{c}-lattice parameter is similar to that of \emph{A}Fe$_2$As$_2$ (\emph{A}=Ba, Sr, Ca, and Eu) and NaFe$_2$As$_2$,\cite{122structure,JohnstonReview, Chu2010} the initial refinement of single crystal x-ray diffraction data was performed with the atomic coordinates of CaFe$_2$As$_2$ but with La and Na at Ca site. Site occupation factors were checked by freeing individual occupancy factors of atom sites. This procedure revealed that the stoichiometry of the crystal is La$_{0.4}$Na$_{0.6}$Fe$_2$As$_2$. The La:Fe:As ratio agrees with that determined from elemental analysis. Table I shows the refinement parameters and selected crystallographic data.

The only difference between the growth performed in this work and that for the growth of LaFeAsO is the use of an Al$_2$O$_3$ crucible inside of Ta tube. With the same starting materials and heat treatment process, the growth always results in La$_{0.4}$Na$_{0.6}$Fe$_2$As$_2$ once the flux is in direct contact with an Al$_2$O$_3$ crucible; but LaFeAsO crystals otherwise. Varying the charge/flux ratio from 1:10 to 1:20 doesn't affect the composition of the as-grown crystals. The above observation indicates that the flux becomes oxygen free once Al$_2$O$_3$ crucible is used to contain the melt. After crystal growth, the Al$_2$O$_3$ crucible turns to be grey and reaction between the crucible and flux can be observed especially along the grain boundaries. The successful growth of LaFeAsO crystals from NaAs flux has been attributed to some reasonable solubility and diffusivity of oxygen in NaAs flux by forming NaAsO$_2$.\cite{YanFluxRequirement} NaAsO$_2$ has a low melting temperature $\sim$600$^\circ$C and can react with the Al$_2$O$_3$ crucible at high homogenizing temperature used in our growth. The reaction might consume the oxygen in the melt and maintain an oxygen-free environment for the growth of La$_{0.4}$Na$_{0.6}$Fe$_2$As$_2$.

\begin{table*}[!ht]
\caption{Selected crystallographic data and refinement parameters for La$_{0.40}$Na$_{0.60}$Fe$_2$As$_2$ at 100\,K and 173\,K.\\}
\label{table1}
\begin{ruledtabular}
\begin{tabular} {lll}
Temperature (K)	& 100(2)& 173(2)\\
Radiation, wavelength (\emph{${\AA}$})&Mo K$\alpha$, 0.71073&Mo K$\alpha$, 0.71073\\
Space group, \emph{Z}	&\emph{Fmmm} (No. 69), 4&\emph{I4/mmm} (No. 139), 2\\
\emph{a} (${\AA}$)	&5.4697(11)	&3.8669(3)\\
\emph{b} (${\AA}$)	&5.4740(11)	&3.8669(3)\\
\emph{c} (${\AA}$)	&12.068(2)&	12.108(2)\\
\emph{V} (${\AA}$$^3$)&	361.34(13)&	181.05(4)	\\
$\theta$ range (º)&	3.38-28.27&	3.37-28.25\\
\emph{R}$_1$* (all data)	&0.0188&	0.0169	\\
\emph{wR}$_2$* (all data)&	0.0487&	0.0410\\
Goodness-of-fit on F$^2$	&1.160&	1.215	\\
Largest diff. peak/hole (e$^-$/${\AA}$$^3$)&	1.64/-3.16&	1.17/-2.04	\\
Atomic parameters	&	& \\
La/Na	&2\emph{a}(0,0,0)	&2\emph{a}(0,0,0) \\
	&U$_{iso}$(${\AA}$$^2$)\,=\,0.00953	& 0.0115	\\
    &Occupancy La/Na\,=\,0.406/0.594	& 0.402/0.598	\\
Fe	&8\emph{f}(0.25,-0.75,0.25)	&4\emph{d}(0,0.5,0.25) \\
	&U$_{iso}$(${\AA}$$^2$)\,=\,0.0078	& 0.0098	 \\
As	&8\emph{i}(0,-0.5,0.13558)	&4\emph{e}(0,0,0.3644) \\
	&U$_{iso}$(${\AA}$$^2$)\,=\,0.00793	& 0.010 \\
Bond lengths	&	& \\
Fe-As (${\AA}$)	&2.3768(4)	& 2.3785(4)\\
Fe-Fe (${\AA}$)	&2.7348(6), 2.7370(6)	& 2.7343(2)\\
Bond angles	&	& \\
As-Fe-As (deg)	&108.97(5),109.69(5),109.76(3)	& 108.76(4),109.83(4)\\

\end{tabular}
\end{ruledtabular}
* \emph{R}$_1$=$\sum$$|$$|$F$_0$$|$-$|$F$_c$$|$$|$/$\sum$$|$F$_0$$|$; \emph{w}\emph{R$_2$}=$|$$\sum$$|$\emph{w}(F$_0$$^2$-F$_c$$^2$)$^2$$|$/$\sum$$|$\emph{w}(F$_0$$^2$)$^2$$|$$|$$^{1/2}$, where \emph{w}=1/$|$$\sigma$$^2$F$_0$$^2$+(AP)$^2$+BP$|$, and P=(F$_0$$^2$+2F$_c$$^2$)/3; A and B are weight coefficients.
\end{table*}

Despite tremendous effort in exploring new iron-based superconductors, there is no report about replacing the alkaline earth ions in \emph{A}Fe$_2$As$_2$ compounds with a mixture of one rare earth ion (\emph{R$^{3+}$}) and one alkali metal, i.e., no compound with the chemical formula  \emph{R$^{3+}$}$_{1-x}$\emph{B}$_x$Fe$_2$As$_2$ (\emph{R}\,=\,rare earth, \emph{B}\,=\,alkali). This is understandable since \emph{R$^{3+}$}Fe$_2$As$_2$ compounds don't exist due to the tolerance of the tetragonal ThCr$_2$Si$_2$ structure to electron count; and NaFe$_2$As$_2$ is a metastable phase, which has been suggested to be due to the size mismatch between the square prismatic site and Na$^+$.\cite{Clarke2010,Chu2010} Thus, both the ionic radius and electron count contribute to the stability of the tetragonal ThCr$_2$Si$_2$ structure.\cite{Hoffmann1985} Considering the little difference in the ionic radii of Na$^+$ (1.18\,${\AA}$), La$^{3+}$ (1.16\,${\AA}$), and Ca$^{2+}$ (1.12\,${\AA}$),\cite{Shannon} the tolerance of the ThCr$_2$Si$_2$ structure to the electron count may play a major role in stabilizing La$_{0.5-x}$Na$_{0.5+x}$Fe$_2$As$_2$ phase. The successful growth of La$_{0.4}$Na$_{0.6}$Fe$_2$As$_2$ single crystals in this work suggests that stable La$_{0.5-x}$Na$_{0.5+x}$Fe$_2$As$_2$ can be synthesized despite the instability of the parent compounds. A wide range of \emph{x} might be expected considering the studies of Ca$_{1-x}$\emph{R}$_x$Fe$_2$As$_2$,\cite{CaPrFe2As2} Ca$_{1-x}$Na$_x$Fe$_2$As$_2$,\cite{CaNaFe2As2} and Ba$_{1-x}$Na$_x$Fe$_2$As$_2$.\cite{BaNaFe2As2}

We noticed that superconductivity was observed in Eu$_{1-x}$\emph{B}$_x$Fe$_2$As$_2$ (\emph{B}=Na, and K), which also has a rare earth ion and alkali metal sharing the 2a site.\cite{EuKFe2As2, EuNaFe2As2} However, Eu is expected to be 2+ and EuFe$_2$As$_2$ has been claimed as a magnetic homologue of SrFe$_2$As$_2$ with the magnetic order of Eu$^{2+}$ sublattice at 20\,K.\cite{EuFe2As2} Varying  the Eu/\emph{B} ratio in Eu$_{1-x}$\emph{B}$_x$Fe$_2$As$_2$ only changes the hole concentration but cannot lead to electron doping.

It has been well established that other \emph{A}Fe$_2$As$_2$ (\emph{A}\,=\,Ca, Sr, Ba, and Eu) members show a structural transition from a high temperature tetragonal phase (space group \emph{I}4\emph{/mmm}) to a low temperature orthorhombic phase (space group \emph{Fmmm}).\cite{JohnstonReview} Single crystal x-ray diffraction performed at 100\,K (see Table I for refinement and structural parameters) confirmed the orthorhombic symmetry for La$_{0.40}$Na$_{0.60}$Fe$_2$As$_2$. This suggests that for La$_{0.40}$Na$_{0.60}$Fe$_2$As$_2$ a structural transition takes place in the temperature range 100\,K$\leq$\,T\,$\leq$173\,K. To further reveal the details of this structural transition, we measured the magnetic, transport, and thermodynamic properties, and performed single crystal neutron diffraction.

\begin{figure}
\centering \includegraphics[width = 0.48\textwidth] {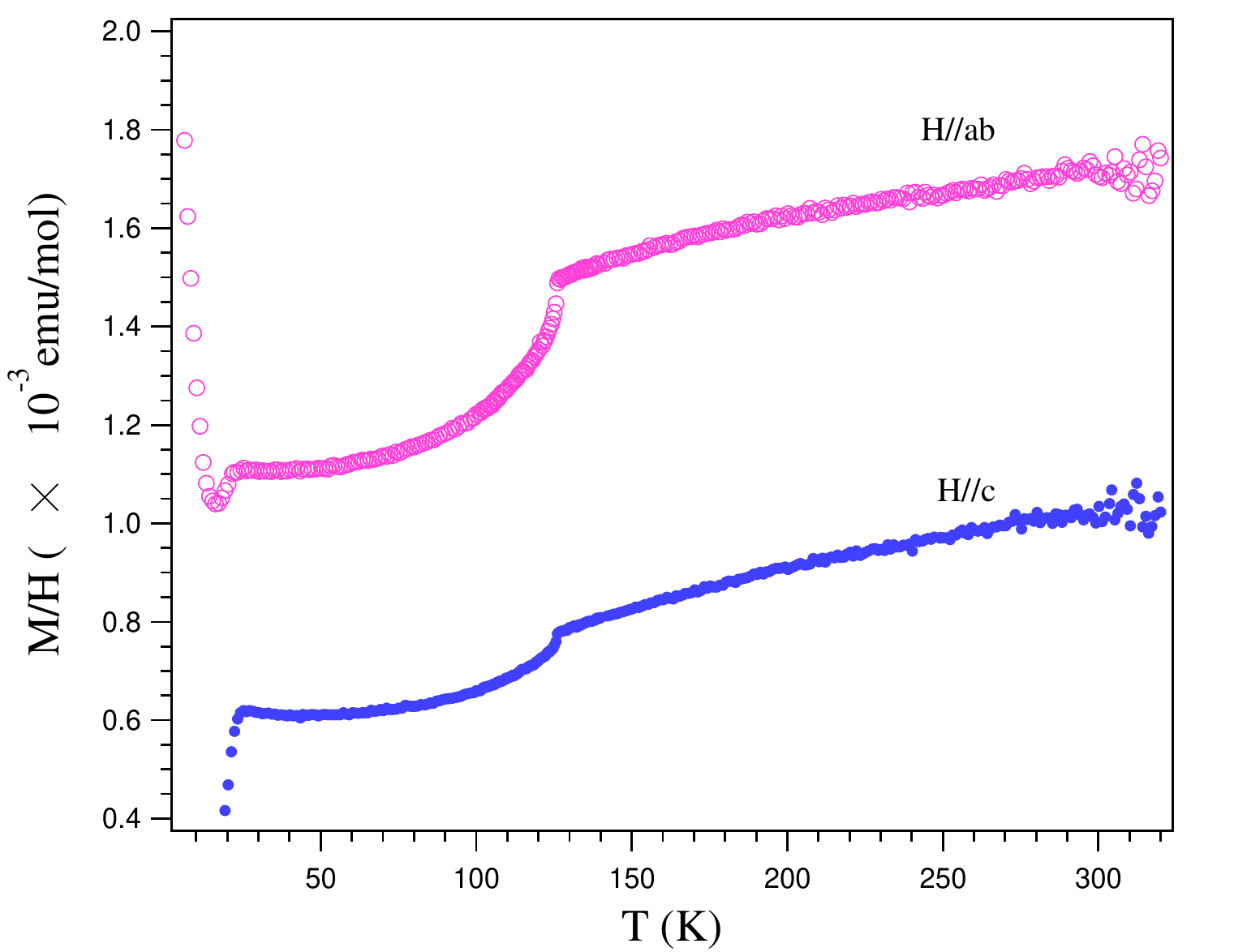}
\caption{(color online) Temperature dependence of magnetic susceptibility measured with a magnetic field of 50\,kOe applied parallel and perpendicular to the plate, respectively.} \label{Mag-1}
\end{figure}

Figure\,\ref{Mag-1} shows the temperature dependence of magnetic susceptibility $\chi$\,=\,\emph{M/H} measured in an applied magnetic field of 50\,kOe. 6 pieces of plate-like crystals were coaligned with \emph{c}-axis parallel to each other. The measurement was performed with the magnetic field applied perpendicular and parallel to the crystallographic \emph{c}-axis, respectively. As shown in Fig.\,\ref{Mag-1}, a clear anisotropy with $\chi$$_{ab}$$>$$\chi$$_{c}$ was observed in the whole temperature range studied, and a steplike change in both $\chi$$_{ab}$ and $\chi$$_{c}$ signals a transition at T$_s$\,=\,125\,K. Above T$_s$, both $\chi$$_{ab}$ and $\chi$$_{c}$ increase linearly with increasing temperature. The temperature dependence and anisotropy of magnetic susceptibility are similar to those of \emph{A}Fe$_2$As$_2$ (A\,=\,Ca, Sr, and Ba). The magnitude of room temperature $\chi$$_{ab}$ and $\chi$$_{c}$ are similar to those of CaFe$_2$As$_2$.\cite{XHChen2008} The anomalies below 28\,K in $\chi$$_{ab}$ and $\chi$$_{c}$ curves are due to surface superconductivity as discussed later.

\begin{figure}
\centering \includegraphics[width = 0.48\textwidth] {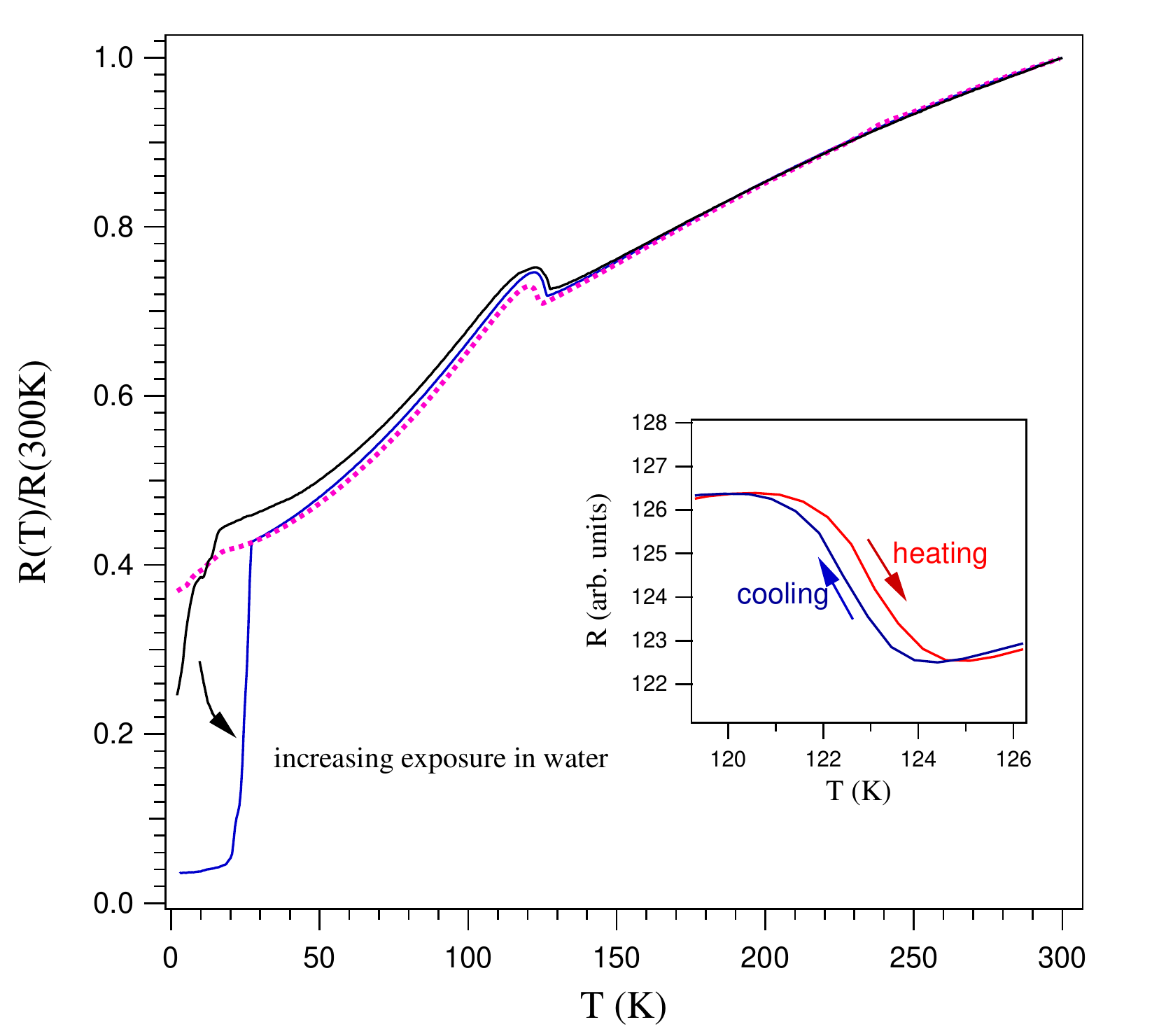}
\caption{(color online) Temperature dependence of normalized in-plane electrical resistivity in zero field. The inset highlights the hysteresis. The dashed curve shows the electrical resistivity after an overnight annealing at 300$^o$C in sealed quartz tube.} \label{RT-1}
\end{figure}

Figure\,\ref{RT-1} shows the temperature dependence of the in-plane electrical resistivity normalized by the room temperature value R(300K). For all pieces measured, an anomaly around 125\,K was observed and this temperature agrees well with that observed in the above magnetic measurement. Across this anomaly, a hysteresis of 0.6\,K, as highlighted in the inset of Fig.\,\ref{RT-1},  was observed suggesting the first order nature of this transition. The increase of electrical resistivity while cooling across 125\,K and the hysteresis are similar to those in CaFe$_2$As$_2$.\cite{XHChen2008} A drop of resistivity was normally observed below 30\,K signaling filamentary superconductivity. This resistivity drop becomes larger in magnitude and sharper in transition width with increasing exposure of crystals in water. Once the crystal was immersed in water for over 6 hours, superconductivity with T$_c$\,$\sim$\,26\,K shows up (Here T$_c$ was defined as the temperature where zero resistivity was observed. Resistivity starts to drop at $\sim$28\,K.).  Heat treatment of the crystals at 300$^o$C in sealed quartz tube can suppress the low temperature drop. As shown in  Figure\,\ref{RT-1}, after an overnight heat treatment, the sharp drop in resistivity is suppressed and shifts to a lower temperature. However, once the heat treated crystals were left in air for further exposure, the suppressed resistivity drop comes back. The recovery of the resistivity drop becomes faster once the crystals were immersed in water. This suggests that the observed superconductivity stays on the crystal surface and is induced by moisture as in SrFe$_2$As$_2$ and FeTe$_{0.8}$S$_{0.2}$.\cite{WaterSrFe2As2, FeTeS} The magnetic measurements (not shown) of one crystal with zero resistivity in a field of 10 Oe also support the surface superconductivity scenario.

\begin{figure}
\centering \includegraphics[width = 0.48\textwidth] {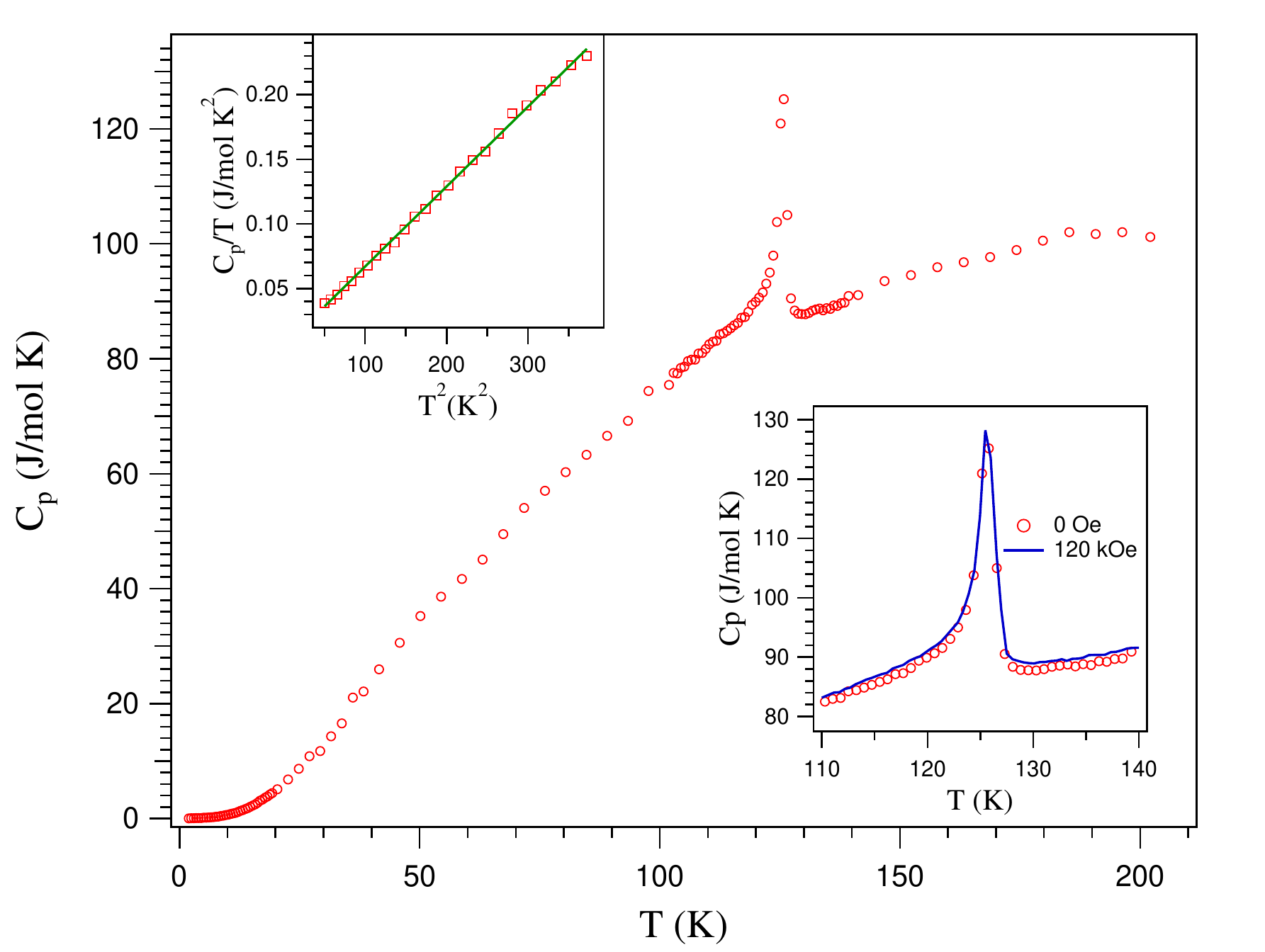}
\caption{(color online) Temperature dependence of specific heat. Upper inset shows the Cp/T as a function of T$^2$ for low temperature data. Lower inset highlights the anomaly around 125\,K in zero and 120\,kOe magnetic fields.} \label{Cp-1}
\end{figure}

\begin{figure*}[t]
\centering \includegraphics[width = 0.80\textwidth] {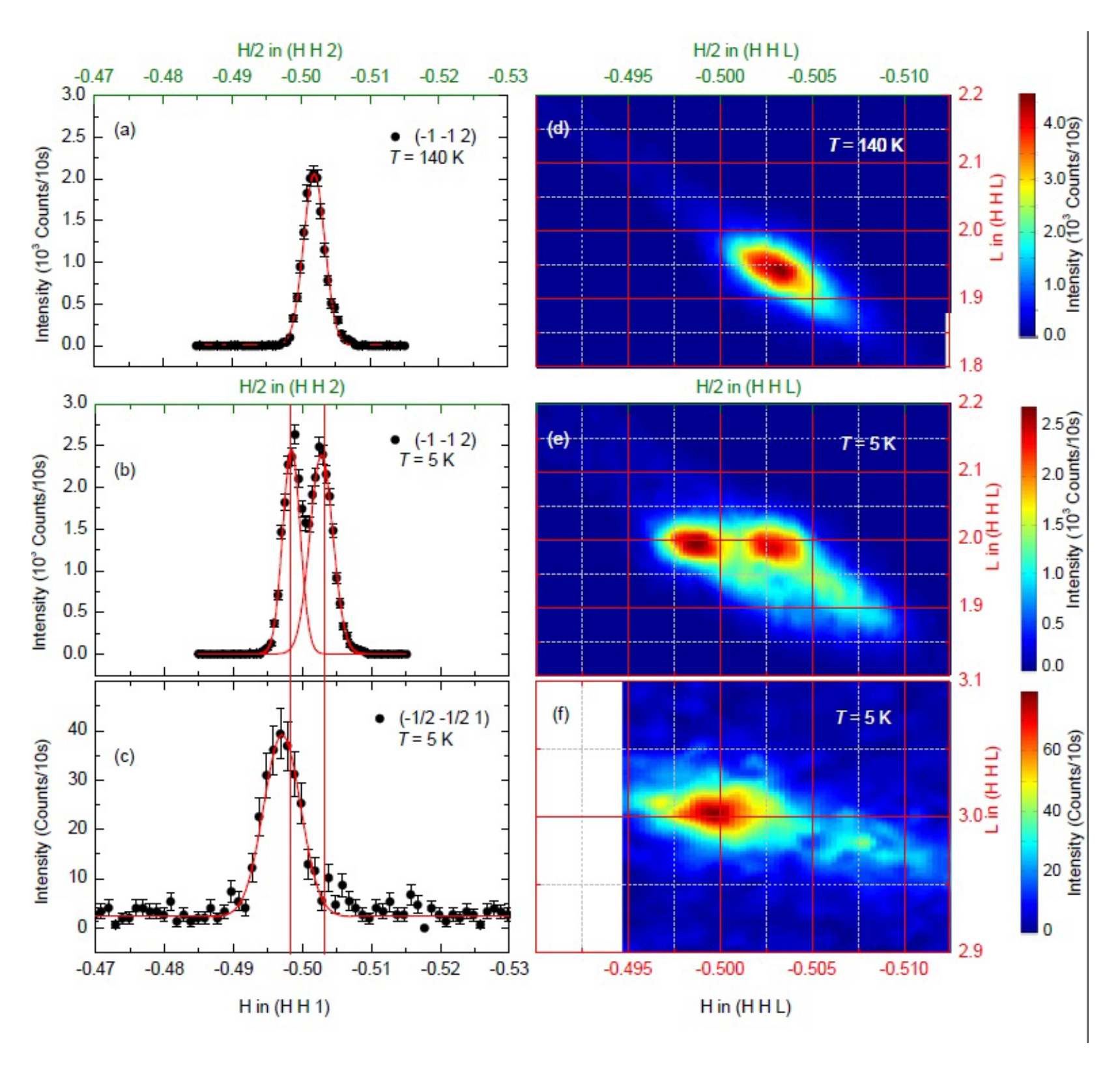}
\caption{(color online)(a) Q scans through the tetragonal (-1 -1 2)$_{T}$
reflection at \emph{T} = 140 K. Only a single peak consistent with
the tetragonal symmetry of the lattice was observed. (b) At low temperature,
the (-1 -1 2)$_{T}$ peak splits into two peaks characterizing the
structural phase transition into the orthorhombic structure. (c) Q
scan through the magnetic (-$\frac{1}{2}$ -$\frac{1}{2}$ 1)$_{T}$
reflection. The magnetic peak is associated with only one structural
twin domain (left) indicating the magnetic propagation vector to be
(1 0 1)$_{O}$. (d-e) Two dimensional maps for the (-1 -1 2)$_{T}$
reflection showing the presence of a single peak at high temperature
and the presence of two equally populated twins at low temperature.
(f) Two dimensional map for the magnetic (-$\frac{1}{2}$ -$\frac{1}{2}$
3)$_{T}$ reflection at \emph{T} = 5 K showing association of the magnetic
peak with the left structural twin. No data were collected in the
white rectangular region.} \label{fig_1-1}
\end{figure*}

Figure\,\ref{Cp-1} shows the temperature dependent specific heat data measured in the temperature interval 1.9\,K\,$\leq$\,T\,$\leq$\,200\,K on one piece of as-grown crystal with a sharp resistivity drop at 26\,K. A pronounced anomaly was observed at T$_s$\,$\approx$\,125\,K. No anomaly was observed around 26\,K, which suggests that the resistivity drop shown in Fig.\,\ref{RT-1} doesn't represent bulk behavior. The lower inset of Fig.\,\ref{Cp-1} shows the data measured in an applied magnetic field of 120\,kOe; little field effect was observed on this lambda-type anomaly. The upper inset to Fig.\,\ref{Cp-1} shows the low-temperature C$_p$/T data plotted as a function of T$^2$.  The fitting of specific heat data in the range 50\,$<$\,T$^2$\,$<$\,350\,K$^2$ to the standard power law, C$_p$/T\,=\,$\gamma$\,$+$\,$\beta$T$^2$ yields $\gamma$\,=\,5.1\,mJ/mol\,K$^2$ and $\beta$\,=\,0.64(1)\,mJ/mol\,K$^4$, where $\gamma$ is the Sommerfeld electronic specific heat coefficient and $\beta$ the coefficient of the Debye T$^3$ lattice heat capacity at low temperatures. The latter gives the Debye temperature $\theta$$_D$ with the following relation $\theta$$_D$\,=\,(12$\pi$$^4$N$_A$k$_B$n/5$\beta$)$^{1/3}$, where n is the number of atoms per formula unit, N$_A$ is Avogadro's constant and k$_B$ is Boltzmann's constant. With n\,=\,10 and $\beta$\,=\,0.64(1)\,mJ/mol\,K$^4$, the Debye temperature is $\theta$$_D$\,=\,250\,K. Both the Sommerfeld coefficient and Debye temperature are comparable to those of CaFe$_2$As$_2$.\cite{NiCaFe2As2}

\begin{figure}
\centering \includegraphics[width = 0.48\textwidth] {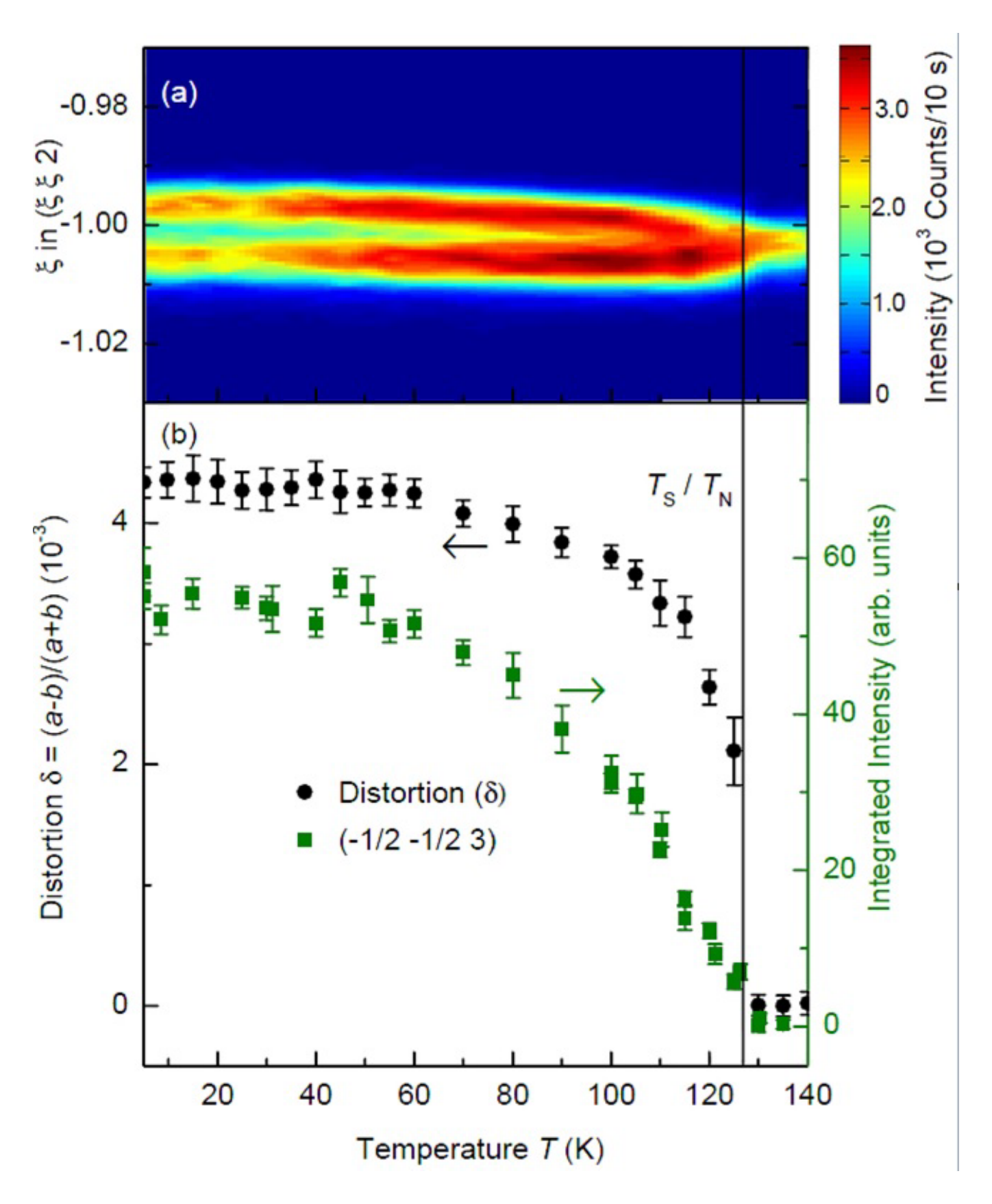}
\caption{(color online) (a) Two-dimensional map of the Q-scans as a function
of temperature for the (-1 -1 2)$_{T}$ reflection. (b) Temperature
dependence of the orthorhombic distortion measured by performing Q
scans through the (-1 -1 2)$_{T}$ reflection. Temperature dependence
of the integrated intensity for the magnetic (-$\frac{1}{2}$ -$\frac{1}{2}$
3)$_{T}$ reflection. Both the temperature dependencies were measured
during heating.} \label{fig_2-1}
\end{figure}

\begin{figure}
\centering \includegraphics[width = 0.48\textwidth] {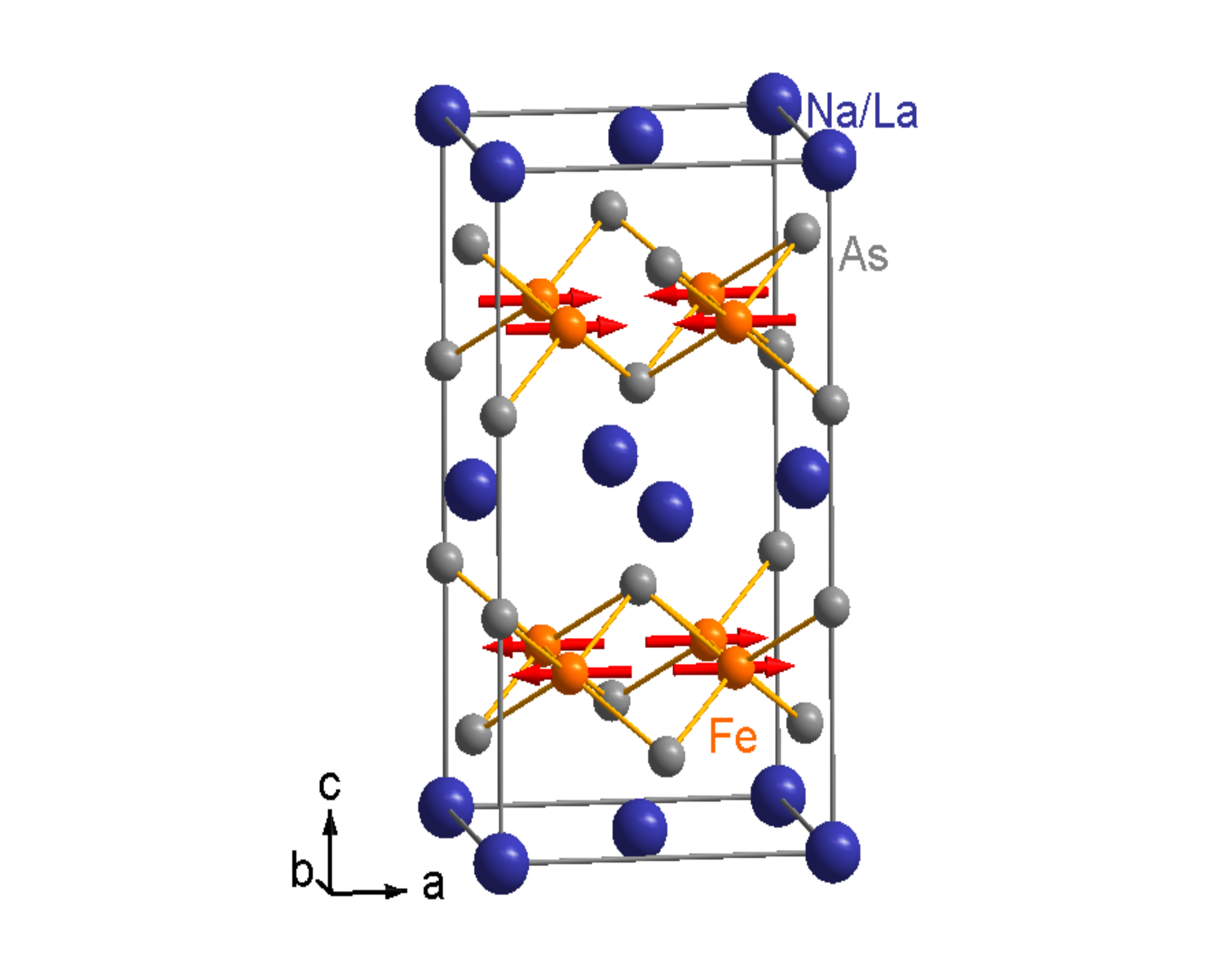}
\caption{(color online) magnetic structure of La$_{0.40}$Na$_{0.60}$Fe$_2$As$_2$.} \label{spin-1}
\end{figure}

To further investigate the  structural phase transition and magnetic order, single crystal neutron diffraction measurements were performed in the temperature range 5\,K$\leq$\,T\,$\leq$150\,K. At \emph{T}\,=\,300 K,
the crystal structure is well described by a tetragonal lattice with
lattice parameters \emph{a} = 3.875(5)${\AA}$ and \emph{c} = 12.224(5) ${\AA}$
. Figure \ref{fig_1-1}\,(a) shows ($\xi$ $\xi$
0) scan through the nuclear (-1 -1 2)$_{T}$ reflection at \emph{T}\,=\,140\,K.
A single peak shape, consistent with the tetragonal symmetry of the
lattice, was observed. Figure \ref{fig_1-1}\,(d) shows corresponding
two-dimensional (2-D) map at the same temperature, further confirming
single peak shape of the (-1 -1 2)$_{T}$ reflection. As the sample
is cooled below \emph{T}$_s$, the (-1 -1 2)$_{T}$
peak splits into two, indicating orthorhombic crystal structure with
$a\neq b$ as shown in Figure \ref{fig_1-1}(b). In fact depending on
the resolution of the instrument and the scattering plane a maximum
of four peaks could be present. Figure \ref{fig_1-1}\,(e)
shows corresponding 2-D map indicating almost equal population of
the twin domains. Having established structural phase transition in
this compound, extensive search for the magnetic peak was performed
at \emph{T} = 5 K. In particular, ($\xi$ $\xi$ 0) scans along the
{[}$\frac{1}{2}$ $\frac{1}{2}$ 0{]}, {[}$\frac{1}{2}$ $\frac{1}{2}$
1{]} and {[}$\frac{1}{2}$ $\frac{1}{2}$ 2{]} directions and \emph{L}-scan
along the {[}0 0 L{]} direction were performed to detect magnetic
peaks, if any. Magnetic signal was found at the (-$\frac{1}{2}$ -$\frac{1}{2}$
1)$_{T}$ position. Figure \ref{fig_1-1}\,(c) shows ($\xi$ $\xi$
0) scan through the magnetic (-$\frac{1}{2}$ -$\frac{1}{2}$ 1)$_{T}$
reflection which is associated with the left nuclear twin in Fig.
\ref{fig_1-1}\,(b). In fact, the \emph{H} position of the magnetic
reflection appears at the exactly half value of the left nuclear twin
which implies that the magnetic reflection is associated with the
larger orthorhombic axis \emph{i.e.} the \emph{a} axis and the magnetic
propagation vector is (1 0 1)$_{O}$. 2D map for the (-$\frac{1}{2}$
-$\frac{1}{2}$ 3)$_{T}$ reflection in Fig.\,\ref{fig_1-1}\,(f) also
confirms that the magnetic propagation vector is (1 0 1)$_{O}$.

To determine the structural phase transition temperature, ($\xi$
$\xi$ 0) scans were performed through the (-1 -1 2)$_{T}$ reflection
as a function of temperature and are shown in Fig. \ref{fig_2-1} (a).
Broadening of the Full-Width-Half-Maxima (FWHM) of the (-1 -1 2)$_{T}$
reflection below \emph{T}$_{\textup{S}}$ = (128.0 $\pm$ 0.5) K indicates
a structural phase transition at this temperature. Figure \ref{fig_2-1}(b)
shows temperature dependence of the orthorhombic distortion, $\delta=\frac{a-b}{a+b}$,
measured during heating of the sample. Orthorhombic distortion increases
continuously to the lowest achievable temperature of 5 K. The orthorhombic distortion is similar to that in BaFe$_{2}$As$_{2}$, \cite{Johrendt2008} but smaller than that in SrFe$_{2}$As$_{2}$ or CaFe$_{2}$As$_{2}$.\cite{NiCaFe2As2, YanSrFe2As2} Figure
\ref{fig_2-1}(b) shows temperature dependence of the magnetic reflection
(-$\frac{1}{2}$ -$\frac{1}{2}$ 3)$_{T}$ indicating onset of the
Fe magnetic order at \emph{T}$_{\textup{N}}$ = (128.0 $\pm$ 0.5)
K, the same temperature as the structural phase transition. At this point it is interesting to compare the results of the structural and magnetic phase transitions in other known 122 compounds. Similar
concomitant structural and magnetic phase transitions were observed
previously in the parent \emph{A}Fe$_{2}$As$_{2}$ (\emph{A} = Ca,
Sr, Ba) compounds.\cite{JohnstonReview,JeffReview} It is likely that the tetragonal to orthorhombic phase transition is absent in the metastable superconductor NaFe$_2$As$_2$. However, no structural studies as a function of temperature are available for the NaFe$_2$As$_2$.\cite{Chu2010, BaNaFe2As2, NaFeAs} Various studies also suggest the absence of the structural phase transition for the other known 122 systems such as KFe$_2$As$_2$, RbFe$_2$As$_2$ and CsFe$_2$As$_2$.\cite{KFe2As2,RbFe2As2,CsFe2As2} For the (Ca$_{1-x}$Na$_x$)Fe$_2$As$_2$ samples, doping dependence studies show the structural phase transition for x$\leq$0.36. From x$\approx$0.36 to the highest studied doping level of x$\approx$0.66, macroscopic measurements show no hint of structural phase transition.\cite{CaNaFe2As2} For the (Ba$_{1-x}$Na$_x$)Fe$_2$As$_2$ samples, the structural and magnetic phase transition temperature decreases monotonically as a function of doping and for x$\geq$0.3, no structural phase transition was observed.\cite{BaNaFe2As2} The magnetic and structural transitions are coincident and first order in (Ba$_{1-x}$Na$_x$)Fe$_2$As$_2$ unlike the Co-doped BaFe$_2$As$_2$ where the two transitions are separated.\cite{Nandi,Ni2010,Canfield2009,Dan,Andy}

We noticed that the transition temperature determined with neutron diffraction is about 3\,K higher than that determined from magnetic susceptibility, electrical resistivity, and specific heat. Since only one transition was observed in all bulk measurements, we believe this 3\,K difference comes from the instrument. Our neutron measurements clearly demonstrate the same  (within 0.5\,K) transition temperature for the structural and magnetic phase transitions.

At low temperatures (\emph{T} = 5 K) a set of structural and magnetic
peaks were collected for the determination of Fe magnetic structure.
The nuclear reflections were fitted with the structural parameters
listed in Table I taking proper account
of the nuclear twins. The magnetic reflections can be well fitted
with the stripe antiferromagnetic structure with moments along the
\emph{a} direction and with an ordered moment of 0.7(1) \,$\mu_{\textup{B}}$
at \emph{T}\,=\,5\,K. Figure\,\ref{spin-1} shows the magnetic structure. According to this magnetic structure, the Fe moments
are antiferromagnetically aligned along the \emph{a} direction and
ferromagnetically along the \emph{b} direction, the same as the reported
stripe antiferromagnetic structure of the \emph{A}Fe$_{2}$As$_{2}$
(\emph{A} = Ca, Sr, Ba) compounds.\cite{JeffReview} A stripe antiferromagnetic structure with moments along the \emph{b}
direction produced much larger $\chi^{2}$ and hence, was discarded.

\section{Conclusions}

We have grown La$_{0.40}$Na$_{0.60}$Fe$_2$As$_2$ single crystals and studied the magnetic and structural transitions. La$_{0.40}$Na$_{0.60}$Fe$_2$As$_2$ shows the following features similar to other well studied \emph{A}Fe$_{2}$As$_{2}$ (\emph{A} = Ca, Sr, Ba) compounds:
(1) It undergoes a structural phase transition at T$_s$\,=125\,K from the high temperature tetragonal phase (space group \emph{I}4\emph{/mmm})to the low temperature orthorhombic phase (space group \emph{Fmmm});
(2) Concomitant with the structural phase transition, the Fe moments (0.7(1)\,$\mu_{\textup{B}}$
at \emph{T}\,=\,5\,K) order along the \emph{a} direction with the low temperature stripe antiferromagnetic structure;
(3)This structural transition is accompanied by anomalies in the temperature dependence of electrical resistivity, anisotropic magnetic susceptibility, and specific heat.

The above similarities suggest that La$_{0.5-x}$Na$_{0.5+x}$Fe$_2$As$_2$, or even compounds with other rare earth (\emph{R}$^{3+}$) and alkali ions in the spacing layer, provides a new material platform for the study of iron-based superconductors. A unique feature for La$_{0.5-x}$Na$_{0.5+x}$Fe$_2$As$_2$ is that it can be tuned from electron-doped (x$<$0) to hole-doped (x$>$0) by varying the La/Na ratio. This enables the study of the electron-hole asymmetry without disturbing the FeAs conducting layer. The composition studied in this work is hole-doped from a simple electron counting, which is supported by our preliminary photoemission study. Following the phase diagrams for the hole-doped Ba$_{1-x}$K$_x$Fe$_2$As$_2$ and the electron-doped BaFe$_{2-x}$Co$_x$As$_2$, where doping suppresses the structural and magnetic transitions, La$_{0.5}$Na$_{0.5}$Fe$_2$As$_2$, which can be looked as one parent compound in La$_{0.5-x}$Na$_{0.5+x}$Fe$_2$As$_2$, would show structural and magnetic transitions at temperatures higher than 125\,K. Our work calls for both theoretical and experimental efforts in this system. The first step is to determine the range of \emph{x} for which La$_{0.5-x}$Na$_{0.5+x}$Fe$_2$As$_2$ is stable.

We noticed that in NaAs flux our growth always obtained La$_{0.40}$Na$_{0.60}$Fe$_2$As$_2$ crystals even though we varied the charge/flux ratio purposely. This signals that La$_{0.40}$Na$_{0.60}$Fe$_2$As$_2$ is the stable phase in NaAs-rich region. Crystal growth in FeAs flux might be a better approach to synthesize compounds with a different La/Na ratio.

\section{Acknowledgments}
JQY thanks Chenglin Zhang, Brandt Jensen, Kevin Dennis, and Alfred Kracher for help in crystal growth and elemental analysis. SN acknowledges S. Mayr  for the technical assistance with the Laue camera. Work at ORNL and at Ames Laboratory was supported by the US Department of Energy, Office of Sciences, Basic Energy Science, Materials Sciences and Engineering Division. Ames Laboratory is operated for the US DOE by Iowa State University under Contract No. DE-AC02-07CH11358.

\end{document}